\documentstyle[12pt,psfig]{article}
\begin{document}

\begin{center}
\large
{\bf EXPLORING THE ULTRAHIGH ENERGY NEUTRINO UNIVERSE}
\end{center}

\vspace{.5in}

\centerline{D.B. Cline\footnote{Physics Dept., UCLA} and F.W. Stecker\footnote{Lab. for High Energy Astrophysics, NASA Goddard Space flight Center}}

\vspace{.5in}


\newpage

\normalsize

\vspace{.5in}

\begin{abstract}
\noindent {\bf Executive Summary:} 
Astronomy at the highest energies observed must
be performed by studying neutrinos rather than photons 
because the universe is opaque to photons at these energies. 
In making observations of neutrinos at energies in
excess of $10^{19}$eV, one can deduce information about the distribution and
time (redshift) history of cosmic rays which may have been accelerated to 
energies above $10^{20}$ eV. Very large fluxes of neutrinos at these energies 
which exhibit a hard spectrum would provide evidence of a phase transition 
predicted by Grand Unified Theories of electromagnetic, weak and strong nuclear
forces which would have occurred as early as $\sim 10^{-35}$ seconds after the 
big-bang. Alternatively, such fluxes may be evidence of superheavy particles
which could make up the mysterious ``dark  matter'' which comprises the bulk 
of the mass of the universe. Such particles are predicted to have been produced
at the end of an inflationary phase of expansion in the very early universe.
Neutrinos at energies around $10^{15}$eV may be produced in observable 
quantities by active galaxies from the acceleration of cosmic rays in the 
vicinities of massive black holes or in relativistic jets produced by these
black holes. Cosmological gamma-ray bursts may also be sources of neutrinos
at these energies. Neutrinos at such energies may oscillate into the specific
species of tau neutrinos which, through regeneration, are capable of 
travelling through the Earth to
produce upward-moving showers of charged particles. At energies in excess of
$10^{21}$ eV, ultrahigh energy cosmic rays can be produced by the
interaction of ultrahigh energy neutrinos with the very cold (2 K) 
neutrinos which are relics of the big-bang. In principle, this phenomenon can
be used to study the mass and spatial distribution of the 2 K big-bang 
neutrinos.
\end{abstract}

\newpage

\section{HIGH ENERGY COSMIC NEUTRINO SCIENCE}

\subsection{Neutrinos from Interactions of Ultrahigh Energy Cosmic Rays
with the 3 K Cosmic Background Radiation}

  Measurements from the COBE (Cosmic Background Explorer) convincingly proved 
that the universe is filled with radiation having the character of a near 
perfect 3 K black body, which is the
cooled down remnant of the hot big bang.  Extremely high energy protons 
(above $\sim 10^{20}$ eV) will
collide with photons of this radiation, producing pions.  Ultrahigh energy neutrinos (greater 
than $\sim 10^{19}$ eV) are
the main result of the decay of these pions.  Using present measurements of the flux of such high energy
protons, it can be shown that measurable numbers of high energy neutrinos can be detected using imaging optics aboard 
satellites looking down at the luminous tracks produced in the atmosphere by showers of
charged particles produced when these neutrinos hit the nuclei of 
atoms in the atmosphere. Such satellite arrays have been proposed with the 
names OWL (Orbiting Wide-angle Light Collectors) and Airwatch.

\subsection{Fossils of Grand Unified Theories and Inflation from the Early Universe}

  The modern scenario for the early history of the big bang, takes account of the work of
particle theorists to unify the forces of nature in the framework of Grand Unified Theories
(GUTs).  This concept extends the very successful work of Nobel Laureates Glashow, Weinberg,
and Salam in unifying the electromagnetic and weak nuclear forces of nature.  In GUTs, these
forces become unified with the strong nuclear force at very high 
energies of $\sim 10^{25}$ eV which occurred only $\sim 10^{-35}$ seconds after the big bang. 
The fossil remnants of this unification are predicted to be very heavy 
``topological defects" in the vacuum of
space. These are localized regions where extremely high densities of mass-energy are trapped. Such defects go by designations such as cosmic strings, monopoles, walls,
necklaces (string bounded by monopoles), and textures, depending on their 
geometrical and topological properties.  Inside a topological defect, the vestiges of the early
universe may be preserved to the present day.  Topological defects are expected to produce very
heavy particles that decay to produce ultrahigh energy neutrinos.  One can learn about
the high energy physics unachievable to collider experiments by studying the neutrino signal from
the topological defects.The annihilation and decay of these
structures and particles is predicted to produce large numbers of neutrinos with energies
approaching the energy of grand unification.  
It has been suggested that this process may
have produced the highest energy cosmic rays yet observed.  The discovery of such large flux
of neutrinos with energies near the GUT energy scale would be {\it prima facie} evidence for grand unification.
The GUT energy scale is over 12 orders of magnitude higher than the 
energies currently available at terrestrial accelerators.  It is 
difficult to imagine that terrestrial accelerators which ever reach 
this energy scale.

Relicts of an early inflationary phase in the history of the universe can also lead to the production of ultrahigh energy neutrinos.
 The homogeneity and flatness of the present universe may imply that a period of very rapid expansion,
called inflation, took place shortly after the big bang.  During inflation, the universe is cold but,
when inflation is over, coherent oscillations of the inflation field reheat it to a high temperature. 
While the inflation is oscillating, a non-thermal production of very heavy particles may take place. 
These particles may survive to the present as a part of dark matter.  Their decays can give origin
to the highest energy cosmic rays, either by emission of hadrons and photons, or through
production of ultrahigh energy neutrinos.  Observation of such ultrahigh energy neutrinos may teach
us about the dark matter of the universe as well as its inflationary history.

\subsection{Neutrinos from Active Galactic Nuclei}

Quasars and other active galactic nuclei (AGN) are most powerful 
continuous emitters of energy in the known universe.
These remarkable objects are fueled by the gravitational energy released by 
matter falling into a supermassive black hole at the center of the quasar
core. The infalling matter accumulates in an accretion disk which heats
up to temperatures high enough to emit large amounts of UV and soft 
X-radiation. The mechanism responsible for the efficient conversion of 
gravitational energy to observed luminous energy in not yet completely 
understood. If this conversion occurs partly through the acceleration of 
particles to relativistic energies, perhaps by the shock formed at the inner 
edge of the accretion disk, then the interactions of the resulting 
high energy cosmic rays with the intense photon fields produced by the disk
at the quasar cores can lead to the copious production of mesons. The 
subsequent decay of these mesons will then produce large fluxes of high energy
neutrinos. Since the gamma-rays and high energy cosmic rays deep in the
intense radiation field of the AGN core will lose their energy rapidly and
not leave the source region, these AGN core sources will only be observable
as high energy neutrino sources.

Radio loud quasars contain jets of plasma streaming out from the vicinity of
the black hole, in many cases with relativistic velocities approaching the
speed of light. In a subcategory of quasars, known as blazars, these jets
are pointed almost directly at us with their observed radiation, from radio 
to gamma-ray wavelengths, beamed toward us. It has been found that most of 
these blazars actually emit the bulk of their energy in the high energy
gamma-ray range. If, as has been suggested, the gamma-radiation from these
objects is the result of interactions of relativistic nuclei, then high energy
neutrinos will be produced with energy fluxes comparable to the gamma-ray
fluxes from these objects. On the other hand, if the blazar gamma-radiation is
produced by purely electromagnetic processes involving only high energy 
electrons, then no neutrino flux will result.
  
\subsection{Neutrinos from Gamma-Ray Bursts} 

Gamma-ray Bursts (GRBs) are nature's most energetic transient phenomenon.  
In a very short time of $\sim$ 0.1 to 100 seconds, these bursts 
can release an energy in gamma-rays alone of the order 
of $10^{52}$ erg.  They are detected at a rate of about a thousand per year by
present instruments. It has been proposed that particles can be accelerated in
these bursts to energies in excess of $10^{20}$ eV, either by shocks
or perhaps by photonically driven plasma waves.

It is now known that most bursts are at cosmological distances corresponding to
moderate redshifts ($z \sim 1$). If cosmic-rays are accelerated in them 
to ultrahigh 
energies, interactions with gamma-rays in the sources leading to the production
of pions has been suggested as a mechanism for producing very high energy 
neutrinos as well. These neutrinos would also arrive at the Earth in a burst
coincident with the gamma-rays.
This is particularly significant since the ultrahigh energy cosmic rays from
moderate redshifts are attenuated by interactions with the 3 K microwave 
radiation from the big-bang and are not expected to reach the Earth themselves
in significant numbers.
This attenuation of cosmic rays of ultrahigh energy is known as the 
Greisen-Zatsepin-Kuzmin (GZK) cutoff.

\subsection{Neutrino Oscillations and Neutrino Observatons}

Recent observations of the disappearance of atmospheric muon-neutrinos
relative to electron neutrinos by the Kamiokande group, and also the zenith 
angle distribution of this effect, may be interpreted as evidence of the 
oscillation of this weakly interacting neutrino state (``flavor'') into 
another neutrino flavor, either tau neutrinos or sterile neutrinos. A 
corollary of such a conclusion is that at least one neutrino state has a 
finite mass. This has very important consequences for our basic theoretical
understanding of the nature of neutrinos and may be evidence for the grand
unification of electromagnetic, weak and strong interactions. 

If muon neutrinos oscillate into tau neutrinos with the 
parameterization implied by the Super-Kamiokande measurements, then the 
fluxes of these two
neutrino flavors observed from astrophysical sources should be equal. This is
because cosmic neutrinos arrive from such large distances that many 
oscillations are expected to occur during their journey, equalizing the fluxes
in both flavor states. 

On the other hand, if these oscillations do not occur, the fluxes of tau 
neutrinos from such sources should be much less than those of muon neutrinos.
This is because muon neutrinos are produced abundantly in the decay of pions
which are easily produced in cosmic sources, whereas tau neutrinos are not.

Thus, by looking for upward moving showers from tau neutrinos, which can 
propagate thorugh the Earth through regeneration at energies above $10^{14}$ 
eV, one can test for the existence of neutrino oscillations.
For large mixing angles between neutrino states, the detection of $10^{14}$
eV tau-neutrino induced upward-moving atmospheric showers from the direction 
of a cosmic source such as an active galaxy or gamma-ray burst at a 
distance of 1 Gpc would occur for a difference of the squares of the mass 
states as small as $\sim 10^{-17}$ eV$^2$, providing an extremely sensitive
test for oscillations between neutrino states with extremely small mass 
differences. 

Another important
signature of ultrahigh energy tau neutrinos is the ``double bang'' which they
would produce. The first shower is produced by the original interaction which
creates a tau particle and a hadronic shower. This is followed by the decay of
the tau which produces the second shower bang. The two bamgs are separated by
a distance of $\sim$ 91.4 $\mu$m times the Lotentz factor of the tau.

\subsection{Z-bursts from Neutrinos and Ultrahigh Energy Cosmic Rays}

\subsubsection{Ultrahigh Energy Cosmic Rays}

  The quest for higher and higher energy cosmic rays goes forward undeterred 
by the expectation
that protons above $10^{20}$ eV from sources farther away than $\sim 100$ Mpc
should strongly depleted by interactions 
with the 3 K 
photons which make up the cosmic background radiation from the big-bang.
Some cosmic rays at these energies have already been detected with 
ground-based observatories such as AGASA (Akeno Giant Air Shower Array) and
Fly's Eye. Satellite observatories such as OWL/Airwatch would increase the 
number
to thousands and may allow the detection of cosmic rays above $\sim 10^{21}$ 
eV, if such cosmic rays exist.  The detection of thousands of events would
enable OWL/Airwatch to obtain an energy spectrum covering an interval where 
different acceleration and topological defect scenarios make different
predictions of the energy spectrum.
 
OWL/Airwatch also has the unmatched capacity to map the arrival directions of
cosmic rays over the entire sky and thus to reveal the locations of strong 
nearby sources and large-scale anisotropies, this owing to
the magnetic stiffness of charged particles of such high energy.  
Thus, OWL/Airwatch can investigate 
energy spectra of any detected sources and also time correlations 
with high energy neutrinos and gamma-rays.

\subsubsection{Z-burst Neutrinos}

An exciting possibility is that the highest energy cosmic-ray events
observed above the GZK cutoff energy provides indirect detection of the relic 
neutrinos predicted in standard big bang cosmology.
The observed thermal 3 K cosmic microwave background (CMB)
permeates the universe as a relic of the big-bang is accompanied by
a 2 K cosmic neutrino background of the same thermal big-bang origin.
It has been proposed that high energy neutrinos interacting within
the GZK attenuation distance with the copious 2 K blackbody neutrinos
and annihilating at the Z-boson resonance energy can produce the
observed ``trans-GZK'' air-shower events.
This is a possible solution to the problem of
how can cosmic rays of ``trans GZK'' energies (above the GZK cutoff), 
which are observed to interact in our atmosphere and produce giant air
showers, get here from the extragalactic sources 
which may be many absorption lengths away, ({\it i.e.}, hundreds of 
megaparsecs). This is because high energy neutrinos, which may originate at 
such large distances throughout the Universe, can reach the Earth or nearby
parts of the Universe and then annihilate to produce ultrahigh energy
cosmic rays. The neutrinos themselves interact too weakly with nuclei in our 
atmosphere to produce the observed high altitude air showers, but they can 
annihilate with CNB neutrinos through the Z-boson resonance to produce 
``local" photons and protons, with a probability of order $10^{-2\pm 1}$.
The resulting Z-boson then decays to produce a shower of hadrons and leptons,
a ``Z-burst''.
The products of the Z decay, as measured at the CERN and SLAC
colliders, include on average 20 photons and 2 nucleons.
These photons and protons are the candidate particles for initiating
the observed trans-GZK air-showers.
Because the annihilation process is resonant, the event energy is unique.
It is $E_{Z-burst} = 4\times 10^{21}[m_{\nu}]^{-1}$~eV
in terms of the neutrino mass (in units of eV).
Each nucleon and photon in the burst carries on average an energy 
which is a few percent of the Z-burst energy.

The Z-burst hypothesis based on the assumption that there exists a
significant flux of neutrinos at $E \sim 10^{22}$~eV, perhaps from topological
defects. Some predictive consequences of this hypothesis are 
(a) that the direction of the
air showers should be close to the directions of their cosmological 
sources,
     (b) that there may be multiple events coming from the directions of the 
strongest sources,
     (c) that there exists a relationship between the 
maximum shower energy attainable and the terrestrially-measured neutrino mass,
 and (d) that there may be an observable large-scale anisotropy caused by a 
clustering of 2 K neutrinos within the cosmic-ray attenuation distance.
\section{NEUTRINO FLUX PREDICTIONS}

\begin{figure}[h]
	
\vspace{-1.5cm}
\begin{flushleft}
\mbox{\psfig{file=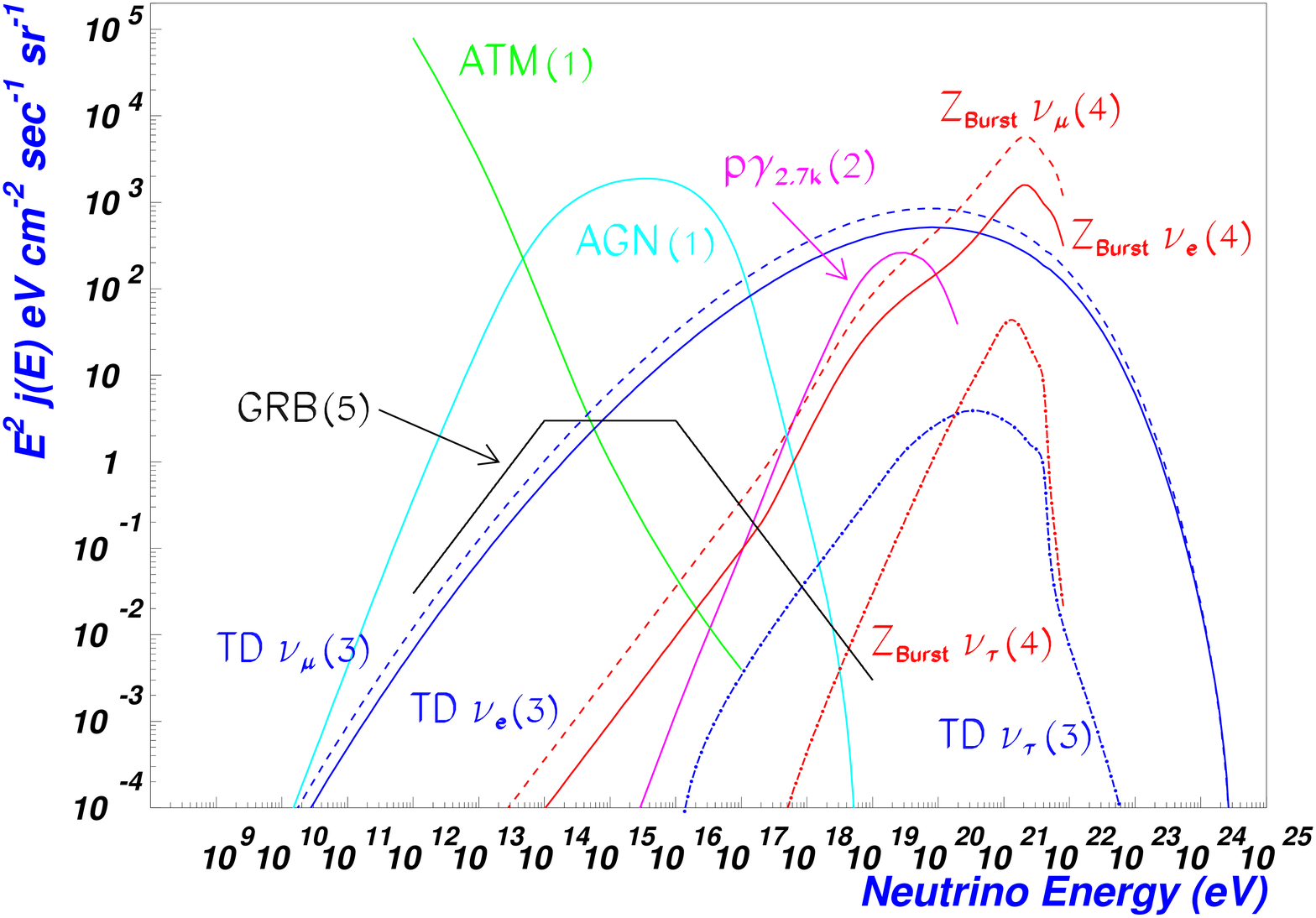,height=5.0in}}
\end{flushleft}
\vspace{-1.cm}
\caption{Neutrino flux predictions: Atmospheric and AGN (1: Stecker 
\& Salamon 1996, Space Sci Rev 75, 341), photomeson production via 
$p\gamma_{2.7K}$ (2: Stecker, Done, Salamon, \& Sommers 1991, 
Phys. Rev. Letters 66, 2697), topological defects (3: Sigl, Lee, 
Bhattacharjee, \& Yoshida 1998, Phys. Rev. D 59, 043504), 
m$_{X} = 10^{16}$ GeV, X $\rightarrow$ q + q,
supersymmetric fragmention), $Z_{Burst}$ (4: Yoshida, Sigl, \& Lee 1998, 
Phys. Rev. Letters  81, 5055), 
m$_{\nu}$ = 1 eV, Primary $\Phi_{\nu} \sim E^{-1}$), and gamma ray bursts (5: 
Waxman \& Bahcall 1997, Phys. Rev. Letters 78, 2292).}
\end{figure}

Figure 1 illustrates the high energy neutrino flux predictions from
various astrophysical sources as a function of neutrino energy.  Note that curves show the differential neutrino flux multiplied by $E_{\nu}^{2}$ which is
equivalent to an energy flux. In the energy range of $10^{14} to 
10^{17}$ eV, the AGN neutrino flux is predicted to dominate over other 
sources. However, neutrinos from individual gamma-ray bursts may be 
observable via their directionality and short, intense time characteristics. 
The time-averaged background flux from all bursts is shown in the figure. 
In the energy range $E \ge 10^{18}$ eV, neutrinos are produced from photomeson
interactions of ultrahigh enrgy cosmic rays with the 3 K background photons.
The highest energy neutrinos ($E \ge 10^{20}$
eV) are presumed to arise from more the speculative physics of topological 
defects and Z-bursts.  

The proposed high energy neutrinos sources also have different signatures in
terms of other observables which include coincidences with other
observations (GRB's), anisotropy (Z-burst's), and specific relations
to the number of hadronic or photonic air showers also induced
by the phenomena (topological defectss, Z-burst's, and 3 K photomeson 
neutrinos).
The distiguishing characteristics of these neutrino
sources are summarized in Table 1.

\begin{table}[h]
	\vspace{0.5cm}
\centering
\begin{tabular}{llllll}
 \hline
 Test & GRB & AGN & TD & Z-Burst & p$\gamma_{2.7K}$ \\
 \hline
 Coincidence & & & & & \\
 with a GRB   & X & - & - & - & - \\
 & & & & & \\
 $(N_{\nu}/N_{p}) \gg 1$& - & - & X & X & X \\
 & & & & & \\
 $(N_{\gamma}/N_{p}) \gg 1$& - & - & X & X & X \\
 & & & & & \\
 Anisotropy & NA & NA & - & X & - \\
 & & & & & \\
 Characterisitic & & & & & \\
 Energy & $10^{14}$ eV & $10^{15}$ eV & $10^{24}$ eV & {\Large $\frac{€10^{20}~{\rm 
 eV}}{m_{\nu} ~({\rm eV})}$} & $10^{19}$ eV \\
 & & & & & \\
 Multiple & & & & & \\
 Events & X & X & - & X & - \\
 \hline 
\end{tabular}
\caption{Distinguishing characteristics of the different sources of 
ultra-high energy neutrinos.}
\end{table}

\section{EXPERIMENTAL SIGNATURES AND RATES}

\subsection{Air Fluorescence Events}

A space-based experiment viewing $10^{6}$ km$^{2}$ of the surface
of the Earth also monitors $\sim 10^{13}$ metric tons of atmosphere.
This large target mass opens the possibility of observing
the interactions of ultra-high energy (UHE) neutrinos in this
atmospheric volume.
The distribution of atmospheric depth of neutrino interactions is 
approximately uniform due to the extremely long interaction path of neutrinos in the atmosphere.
This offers a unique signature of neutrino-induced airshowers as a 
significant portion of the neutrino interactions will be deep in the 
atmosphere, i.e. near horizontal, and well separated from airshowers induced by hadrons
and photons.  
A space-based experiment observing
the fluorescence signal of airshowers will have a segmented
detector plane (pixels) in order to measure the spatial development
of the airshowers.  Additionally, the signals will be recorded with a very
fine time resolution in order to measure the temporal development of
the showers.  This experimental configuration yields a multiple
pixel signature for near horizontal airshowers which translates
into straight-forward  detection
with sufficient angular resolution to
guarantee separation
of neutrino induced events from the hadronic (or electromagnetic) airshower
background.

At ultrahigh energies, the cross sections for neutrino and antineutrino
interactions with quarks become equivalent
(Ghandhi, {\it et al.} 1998, Phys Rev D, 58, 093009).
The kinematics of UHE neutrino 
interactions ($E_{\nu} > 10^{9}$ GeV) leads to the condition that
average energy of the resulting lepton will be approximately 80\% of
the incident neutrino energy.  The remaining 20\% will materialize
in the form of a hadronic cascade from the neutrino interaction point.
Charged current neutrino interactions will, on average, yield an UHE 
charged
lepton and a hadronic airshower.
At these energies,
electrons will generate electronic airshowers while
muons and taus will make airshowers with reduced
particle densities and, thus, fluorescence signals.  The taus offer an additional,
``double-bang'' signature
because of their lifetime and decay modes.  At $10^{10}$ GeV,
$\gamma c \tau = 500$ km for a tau after which the particle will
decay inducing a second airshower separated from the first, hadronic
airshower at the neutrino interaction point.

As a first step in quantifing the rates of neutrino airshower 
observation, we will focus on the charged current electron neutrino interaction.
These deposit 100\% of the incident
neutrino energy into an airshower and will yield the highest
air fluorescence signal for a given neutrino energy
of the possible flavor channels.  Preliminary Monte Carlo
simulation of an OWL/AirWatch instrument have
indicated that charged current electron neutrino interactions
can be identified with a
neutrino aperture
of 20 km$^{2}$-ster at a threshold
energy of $3\times 10^{10}$ GeV and this aperture
grows  with the $E_{\nu}^{0.363}$ assumed increase in neutrino
cross section.   Event rates can be obtained by convolving
this neutrino aperture with neutrino flux predictions and
integrating.  Note that the neutrino interaction
cross section is included in the definition of neutrino aperture.
Assuming a 10\% duty cycle of the experiment,
Table 2 lists the electron neutrino event rates from several
possible UHE neutrino sources: neutrinos from the interaction
of UHE protons with the microwave background (p$\gamma_{2.7K}$,
Stecker {\it et al.} 1991, Phys. Rev. Letters 66, 2697)
and the more speculative sources of topological defects
(Sigl {\it et al.} 1999, Phys. Rev. D 59, 043504) and
the interaction of UHE massive neutrinos with the 2 K relic
neutrino background (Z$_{\rm Burst}$,  Yoshida {\it et al.} 1998, 
Phys. Rev. Letters 81, 5505).

\begin{table}[h]
\centering
\begin{tabular}{|l|c|c|c|}
 \hline
 & p$\gamma_{2.7K}$ & Topological Defects & Z$_{\rm Burst}$ \\
 \hline
 $\nu_{e}$ Events/Year & 5 & 16 & 9 \\
 \hline 
\end{tabular}
\caption{Anticipated electron neutrino, charged current event rates in
a baseline OWL/AirWatch experiment.}
\end{table}

\subsection{Upward Cherenkov Events}

\indent The ensemble of charged particles in an airshower will produce a large
photon signal from Cherenkov radiation which is strongly peaked in the forward
direction and which is much stronger than the signal
due to air fluorescence at a given energy.  This
translates into a much reduced energy threshold for observing
airshowers {\it via} Cherenkov radiation.  As this signal is
highly directional, an orbitting instrument will
only observe those events where the airshower is moving towards
the experiment with the instrument located in the field of the
narrow, Cherenkov cone.

Virtually all 
particles, including neutrinos with $E \ge 40$ TeV, are attenuated by the 
Earth. However, tau neutrinos will regenerate themselves, albeit at a
lower energy, due to the
fact that both charged and neutral current interactions will have
a tau neutrino in the eventual, final state (see section 1.5).  Thus, 
the use of the Earth as a tau neutrino filter leads to the possibility
of a cosmological, long-baseline tau neutrino appearance 
experiment. Moreover, an experiment which monitors
$\sim 10^{6}$ km$^{2}$ of the Earth's 
surface samples an incredible target mass of
the Earth's crust.

At high energies ($E > 10^{6}$ GeV), neutrinos and antineutrinos 
interact with approximately equal cross sections.  Furthermore, the
average energy in the lepton resulting from a neutrino interaction
is greater than 70\% of the incident neutrino energy.  Tau leptons
produced in charged current, neutrino interactions will have a
flight path of length $\gamma c \tau ~(\approx 50~{\rm m}$ at $10^{6}$ 
GeV), after which they will decay producing high energy cascades
for most of the branching ratio.  For 
interactions in the Earth's crust, those events which occur
at a depth less than $\gamma c \tau$ will have a tau coming
out of the Earth and generating an airshower.  For a target
area of 10$^{6}$ km$^{2}$, this yields a target mass
of $10^{8} \times (E_{\nu}/{\rm GeV}$) metric tons, e.g.
$10^{14}$ metric
tons at an energy of $10^{6}$ GeV.

Preliminary investigation of the response of an OWL/AirWatch 
instrument has indicated that the experiment would have
a threshold energy $\ge 10^{5}$ GeV to upward, Cherenkov airshowers.
Assuming that the Super-Kamiokande atmospheric neutrino
results are due to
$\nu_{\mu} \rightarrow \nu_{\tau}$ oscillations,  the 
predicted AGN muon neutrino flux (Stecker \& Salamon 1996, Space Sci. Rev. 
75, 341)
indicates that OWL/AirWatch could observe several hundred tau
events per year.  Thus OWL/AirWatch would measure the flux
of AGN neutrinos and observe their oscillations.

\section{Conclusion}

An OWL/Airwatch array of satellite-based optics optimized to
monitor a significant volume of Earth's atmosphere for the purpose of
studying the characteristics of giant atmospheric air showers induced by 
ultrahigh energy neutrinos (as well as ultrahigh energy cosmic rays) can open
a new window of astronomy and physics which will explore the ultrahigh energy
neutrino universe.

\section{Acknowledgments}

This white paper is the result of a workshop on Observing Ultrahigh Energy 
Neutrinos with OWL/Airwatch which was held at UCLA on November 1-3, 1999.
Contributers to this paper were (in alphabetical order) D.B. Cline, 
G. Gelmini, J. Krizmanic, A. Kusenko, J. Linsley, F.W. Stecker, Y. 
Takahashi and T. Weiler.



\end{document}